\documentclass[aps,pra,twocolumn,eqsecnum,superscriptaddress,showpacs,floatfix,nofootinbib]{revtex4-1}
\usepackage{graphicx}
\usepackage{bm}
\usepackage{dsfont}
\usepackage{ulem}
\usepackage{amsmath,amsfonts,amssymb}
\usepackage{float}
\usepackage{hyperref}
\usepackage{amsmath}
\usepackage{amsfonts}
\usepackage[dvipsnames,usenames]{color}
\usepackage{amssymb}

\hypersetup{colorlinks=true,linkcolor=blue,citecolor=blue,urlcolor=blue}

\input{tcilatex}

\begin{document}

\title{Unusual behavior of sound velocity of a Bose gas in an optical superlattice at quasi-one-dimension}
\author{Lei Chen}
\affiliation{Shenyang National Laboratory for Materials Science,
Institute of Metal Research, Chinese Academy of Sciences, Wenhua
Road 72, Shenyang 110016, China}
\author{Zhu Chen}
\affiliation{National Key Laboratory of Science and Technology on Computational Physics, Institute of Applied Physics and Computational Mathematics, Beijing 100088, China}
\author{Wu Li}
\affiliation{Shenyang National Laboratory for Materials Science,
Institute of Metal Research, Chinese Academy of Sciences, Wenhua
Road 72, Shenyang 110016, China}
\author{Zhidong Zhang}
\affiliation{Shenyang National Laboratory for Materials Science,
Institute of Metal Research, Chinese Academy of Sciences, Wenhua
Road 72, Shenyang 110016, China}
\author{Zhaoxin Liang }
\email{Corresponding author: zhxliang@gmail.com}
\affiliation{Shenyang National Laboratory for Materials Science,
Institute of Metal Research, Chinese Academy of Sciences, Wenhua
Road 72, Shenyang 110016, China}
\date{\today}

\begin{abstract}
A Bose gas trapped in a one-dimensional optical superlattice has emerged as a novel superfluid
characterized by tunable lattice topologies and tailored band structures. In this work, we focus on the
propagation of sound in such a novel system and have found new features on sound velocity, which arises
from the interplay between the two lattices with different periodicity and is not present in the case of a
condensate in a monochromatic optical lattice. Particularly, this is the first time that the sound velocity
is found to first increase and then decrease as the superlattice strength increases even at one dimension.
Such unusual behavior can be analytically understood in terms of the competition between the decreasing
compressibility and the increasing effective mass due to the increasing superlattice strength. This result
suggests a new route to engineer the sound velocity by manipulating the superlattice's parameters. All the
calculations based on the mean-field theory are justified by checking the exponent $\gamma$ of the off-diagonal
one-body density matrix that is much smaller than 1. Finally, the conditions for possible experimental
realization of our scenario are also discussed.
\end{abstract}

\pacs{67.85.-d,67.85.De,67.85.Hj,03.75.Kk}
\maketitle

\section{Introduction}

Sound propagation plays a fundamental role in understanding the superfluid
behavior \cite{1,2,3,4,5,6}. Ever since
the first achievement of Bose-Einstein condensate (BEC) in atomic gases, the
sound velocity has been one of the first things to be studied theoretically
\cite{1} and experimentally \cite{7,8,9,10,11} on a BEC in the
presence of harmonic traps \cite{12,13,14,15}, optical lattices \cite{16,17,18,19,20,21,22,23,24,25,26,27}
and disorder \cite{28,29}, etc. More recently,  the studies on the
sound velocity have been renewed, which are much more in line with its
important applications in quantum simulations involving superfluid \cite{30,31,32}. For instance, in recent proposals concerning the
preparations of many-body states and non-equilibrium quantum phases based on
engineering a superfluid reservoir \cite{33,34,35,36},
the sound velocity acts as a key parameter in determining the system-bath
coupling. The control of sound velocity of a superfluid, such as by using
various carefully configured external traps \cite{16,17,18,19,20,21,22,23,24,25,26,27,28,29}
, therefore constitutes an important ingredient of the reservoir engineering.

In this work, we study the sound velocity of a BEC loaded in an optical
superlattice (OSL). The motivation behind this work ties closely to the
recent progress in engineering novel optical lattices in atomic setups,
which is highlighted by the advances in the superlattice technology.
Compared with conventional monochromatic optical lattice (OL), an OSL is
characterized by several distinguishing features \cite{37,38,39,40,41,42} that we have: (i) an
additional OL of $d_{2}$ periodicity superimposed on a fundamental OL of $%
d_{1}$ periodicity, and (ii) a complete control of the relative phase
between two lattices. In the early days of OSL experiments, the most of
investigations have either focused on the static and dynamical properties of
the condensate like the coherence \cite{38,43} or employed two
non-commensurate lattices to simulate disorder quantum system \cite{44}. More recently, emphasis has been shifted to a BEC loaded in an OSL that
has emerged as a novel kind of superfluid with tunable real-space lattice
topologies (sub-structures within the unit cell) and band structures \cite{37}. In more details, loading cold atoms in an OSL with configurable
structures, together with non-equilibrium control of lattice intensity and
phases, has led to, for example, the observation of the Zak phase connected
to the topology of the band structure at one dimension (1D) \cite{45,46}, as well as opened a new avenue in atomic implementations \cite{47,48}, such as
the simulations of Dirac fermions in interacting relativistic field
theories \cite{49}, and the controlled coherent transport of atomic wave packets \cite{50} or
charge pumping \cite{51}. Along this research line, we expect that an OSL can lead to
new features on the sound velocity not present in the conventional
monochromatic OL.

To further motivate our investigation on the sound velocity of a BEC in an
OSL, we first recall that, in an OL, the behavior of sound velocity of a BEC
is believed to be determined by the interplay among three parameters: the
strength of the optical lattice, $V_{1}$; the interaction between atoms, $c$%
; and the lattice dimension, D (D=1, 2, and 3). Theoretical studies have
shown: (i) when D=1, the sound velocity always decreases monotonically
with increasing $V_{1}$ \cite{18,19,20,21,22,23,24}. (ii) for D=2 and 3, when $c$ exceeds a critical
value, the sound speed first increases to a maximum value and then decreases
with increasing $V_{1}$ \cite{25,26}. When D=3, the sound velocity can even oscillate
with respect to $V_{1}$ \cite{26}. Such rich behavior of sound velocity in an OL can
be understood in terms of compressibility $\kappa $ and effective mass $%
m^{\ast }$ as $c_{s}=\sqrt{1/\kappa m^{\ast }}$. So far all the investigations
suggest that the sound velocity of an optically-trapped 1D BEC
always decreases monotonically with increasing the lattice strength.
Considering that both $m^{\ast }$ and $\kappa $ may be strongly affected by
the topological structure of a 1D OSL due to the introduction of
additional freedom, we expect that a different behavior of sound velocity of
a BEC in an OSL may occur.

In this paper, we are then motivated to launch systematic studies on an
interacting BEC loaded into an OSL at quasi-1D using both the
analytical and numerical methods and show that these systems display the
nontrivial properties of the sound velocity, which share the same physical
origins of sound velocity in 2D (3D) BEC in terms of $%
c_{s}=\sqrt{1/\kappa m^{\ast }}$. For example, our analytical results in the limit
of weak potential display that both the relative phase and the strength of
the two lattices that form a 1D OSL can significantly influence
the sound velocity of a BEC. In particular, a surprise indeed arises that
the sound velocity can first increase and then decrease when the OSL
strength increases even at 1D, which presents a contrast to a
quasi-1D BEC in a 1D monochromatic OL where the sound velocity
always decrease with increasing lattice strength. Such behavior of sound
velocity are then verified by the numerical results and can be understood in
terms of the competition between the compressibility $\kappa $ and the
effective mass $m^{\ast }$. In addition, our further numerical results show
that the sound velocity exhibits very rich behavior with respect to various
choices of OSL parameters. Our study suggests a new route to engineer the
sound velocity by loading a superfluid in an OSL.

The paper is organized as follows. In Section \ref{sec:Model} we derive the
effective model for a quasi-1D BEC in an OSL. In Section~\ref%
{sec:Methods} we study the sound propagation and its velocity of the model
system, using both analytical and numerical approaches, in different
parameter regimes, and finally in Section~\ref{sec:Summary} we summarize our
results and give an outlook.

\section{A quasi-1D BEC in optical superlattice potentials}
\label{sec:Model}
\subsection{Effective model}
\begin{figure}[tbp]
\includegraphics[width=0.46\textwidth]{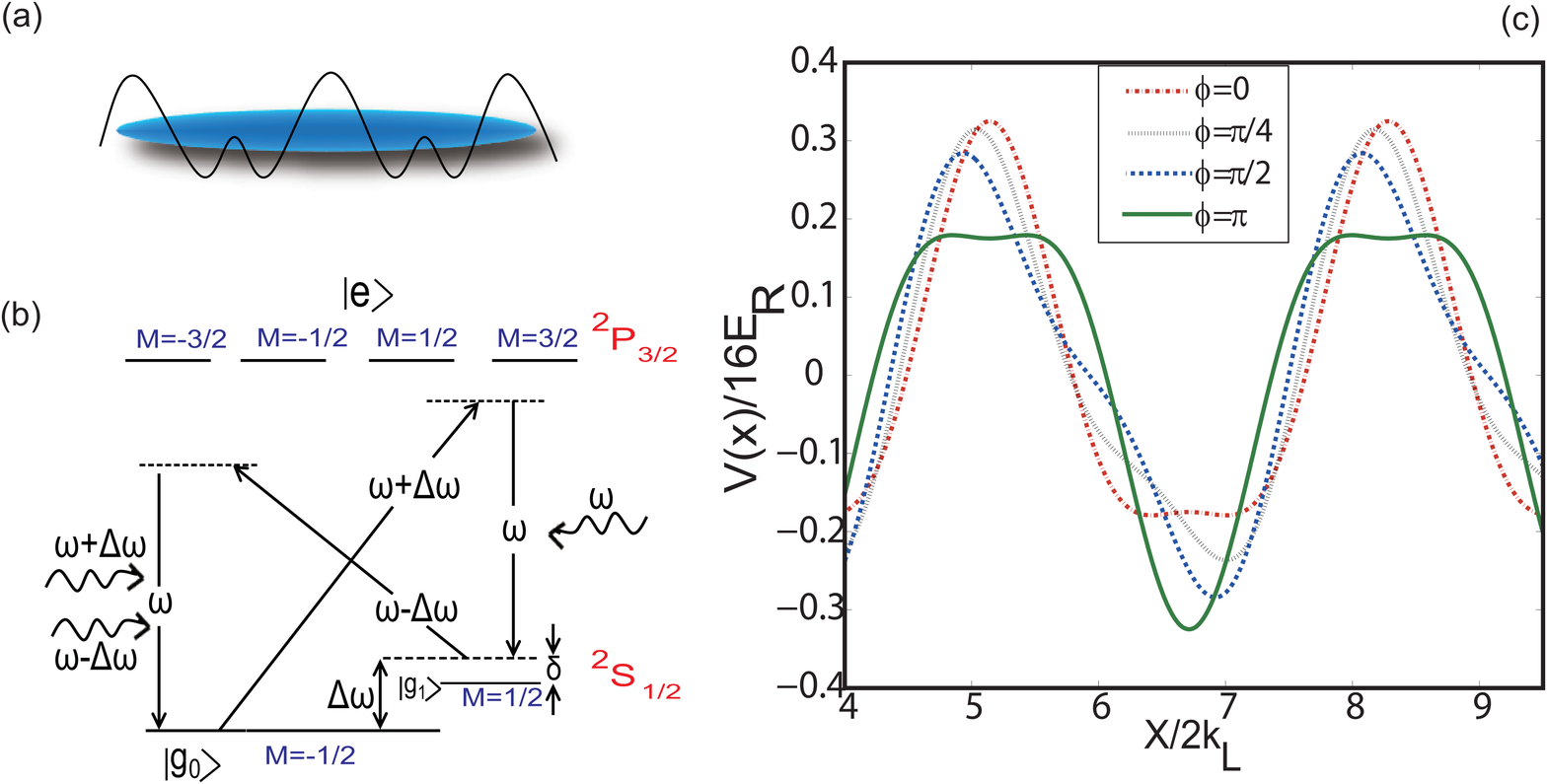}
\caption{(Color online) (a) Quasi-1D BEC trapped in an optical superlattice.
(b) A four-photon optical superlattice for $^{87}\mathrm{Rb}$ atoms: the two states $\left|F=1,M_{F}=-1\right\rangle $
and $\left|F=1,M_{F}=0\right\rangle $ serves as $\left|g_{0}\right\rangle $
and $\left|g_{1}\right\rangle $ and the $5P_{3/2}$ excited state
manifold as $\left|e\right\rangle $. The supperlattice depth $V_{2}$
is given by $\left(\hbar/2\delta\right)\Omega_{\mathrm{eff}}^{+}\Omega_{\mathrm{eff}}^{-}$
with $\Omega_{\mathrm{eff}}^{+}=\Omega_{\mathrm{g_{0},e}}^{+}\Omega_{\mathrm{e,g_{1}}}^{0}/2\Delta$
and $\Omega_{\mathrm{eff}}^{-}=\Omega_{\mathrm{g_{1},e}}^{-}\Omega_{\mathrm{e,g_{0}}}^{0}/2\Delta$.
Here, $\Omega_{\mathrm{g_{0},e}}^{+}$ and $\Omega_{\mathrm{e,g_{0}}}^{0}$
are the Rabi frequencies for the transitions $\left|g_{0}\right\rangle \leftrightarrow\left|e\right\rangle $
and $\left|e\right\rangle \leftrightarrow\left|g_{1}\right\rangle $
driven by the light fields with frequency $\omega+\Delta\omega$ and
$\omega$, respectively \cite{39,40,41,42}. Furthermore, we assume that the condition
of $\Delta\gg\Delta\omega$ can be fulfilled for all three optical
frequencies. (c) Typical optical superlattice structure with different
values $\phi=2\theta+\pi$. The parameters are given as $V_{1}=v_{1}\times16E_{R}=4E_{R}$
and $V_{2}=v_{1}\times E_{R}=1.2E_{R}$, respectively. }
\label{Fig:1}
\end{figure}

We consider a bulk BEC (see Fig. \ref{Fig:1}a) trapped in an OSL of $V_{\text{OSL}}(x)$
along the $x$-direction, whereas the model system is uniform in the $y$- and
the $z$-directions \cite{52}. The OSL of $V_{\text{OSL}}(x)$ is made of a fundamental
lattice in the $x$-direction with the spatial period $d$ denoted as the
primary lattice, and an additional lattice with a period $d/2$ named by the
secondary lattice. The corresponding expression for $V_{\text{OSL}}(x)$ is
\begin{equation}
V_{\text{OSL}}\left( x\right) =V_{1}\sin ^{2}\left( k_{L}x\right) +V_{2}\sin
^{2}\left( 2k_{L}x+\theta \right) .  \label{eq1}
\end{equation}%
Here, $k_{L}=\pi /d$ ($\lambda =2d$) is the wave vector (wave length) of the
laser light creating the OSL and $\theta $ is the relative phase between the
two constituting lattices characterized by the lattice strength $V_{1(2)}$
in unit of the recoil energy of $E_{R}=\hbar ^{2}k_{L}^{2}/2m$. Such an OSL
in equation (\ref{eq1}) has been experimentally realized for quantum gases in
references \cite{39,40} with $V_{1(2)}$ depending on the internal state of an atom (see
Fig. \ref{Fig:1}b for a typical scheme of OSL in bosonic atomic gases). For the
condensate density along the transverse directions in uniform, the freedom
along $y$- and $z$-directions decouples from the $x$-direction, leading to
the realization of a quasi-1D geometry \cite{1,2,26},
\begin{equation}
H-\mu N=\frac{1}{d}\int dx\Psi ^{\ast }\Big[-\frac{\hbar ^{2}}{2m}\frac{%
\partial ^{2}}{\partial x^{2}}+V_{\text{OSL}}(x)+\frac{gn_{0}d}{2}|\Psi |^{2}%
\Big]\Psi ,  \label{eq2}
\end{equation}%
where $\Psi \left( x\right) $ is the condensate wave function satisfying the
normalizing condition $1=\int_{0}^{d}dx\Psi ^{\ast }(x)\Psi \left( x\right) $
with $m$ being the atomic mass, $\mu $ the chemical potential, $N$ the atom
number in condensate, $n_{0}$ the 3D average density and $gn_{0}$ capturing
the role of interactions in the system. In general, the presence of external
confinements will affect the collision property between two particles and
the $g$ can deviate substantially from the coupling constant $g_{3D}$ of a
free 3D Bose gas ( $g_{3D}=4\pi \hbar ^{2}a_{3D}/m$ with $a_{3D}$ being the
3D scattering length). The derivation of $g$ in the presence of external
potentials is highly nontrivial \cite{53}. Here, following reference
\cite{19}, we limit ourself to the parameter regime $g=a_{3D}$ and the
typical values of $gn_{0}d/E_{R}$ in experiments ranges from 0.02 to 1.

Next, we rescale equation (\ref{eq2}) by introducing dimensionless variables, $%
x^{\prime}=2k_{L}x$, $t^{\prime }=t/8E_{R}$, $\psi =\Psi /\sqrt{n_{0}}
$, $\phi =2\theta +\pi $, $v_{1\left( 2\right) }=V_{1\left( 2\right)
}/16E_{R}$, $c=gn_{0}d/8E_{R}$, and $\mu /8E_{R}\rightarrow \mu $. Finally,
we arrive at an effective 1D Hamiltonian describing our model system \cite{54,55,56,57},
\begin{eqnarray}
H_{1D}=\frac{1}{2\pi }\int_{-\pi }^{\pi }dx\psi ^{\ast }\left(
x\right) \Big[ &-&\frac{1}{2}\frac{\partial ^{2}}{\partial x^{2}}+V_{\text{%
OSL}}\left( x\right)   \notag \\
&+&\frac{c}{2}\left\vert \psi \left( x\right) \right\vert ^{2}-\mu \Big]\psi
\left( x\right) ,  \label{eq3}
\end{eqnarray}%
with
\begin{equation}
V_{\text{OSL}}\left( x\right) =V_{1}\cos \left( x\right) +V_{2}\cos \left(
2x+\theta \right) .  \label{eq4}
\end{equation}%
The Hamiltonian (\ref{eq3}) describes a quasi-1D BEC in an OSL at the
mean-field level. The corresponding physics is determined by the effective
interatomic interaction $c$, and the parameters characterizing an OSL of $V_{%
\text{OSL}}\left( x\right) $ ($V_{1(2)}$ and $\phi $). In order to visualize
the superlattice potential $V_{\text{OSL}}(x)$ and highlight the role of the
relative phase $\phi $ in affecting the system, we have plotted $V_{\text{OSL%
}}(x)$ (see Fig. \ref{Fig:1}c) as a function of the relative phase $\phi $.
It is obvious from the plot that, when $\phi $ increases from $0$ to $\pi $
with fixed lattice intensities $v_{1(2)}$, the OSL potential becomes more
squeezed, which can result in enhanced repulsive interatomic interaction.

The quasi-1D condensate wave function $\psi (x)$ and the chemical potential $%
\mu $ in equation (\ref{eq3}) are determined from the corresponding
Gross-Piteavskii equation (GPE) reading \cite{1},
\begin{equation}
\left[ -\frac{1}{2}\frac{\partial ^{2}}{\partial x^{2}}+V_{\text{OSL}}\left(
x\right) +c\left\vert \psi (x)\right\vert ^{2}\right] \psi \left( x\right)
=\mu \psi \left( x\right) .  \label{eq5}
\end{equation}%
Note that for OSL in equation (\ref{eq4}), the Hamiltonian (\ref{eq3}) is still
periodic in the $x$-direction, so the condensate function $\psi (x)$ is
represented by a Bloch state satisfying
\begin{equation}
\psi (x)=e^{ikx}\varphi _{k}(x),  \label{eq6}
\end{equation}%
with
\begin{equation}
\varphi _{k}\left( x\right) =\sum_{m}a_{m}e^{imx}=\sum_{m}(|a_{m}|e^{i\phi
_{m}})e^{imx},  \label{eq7}
\end{equation}%
where $k$ is the Bloch wave vector, $\varphi _{k}$ is a periodic function
with the periodicity of $2\pi $, and $a_{m}$ is the expansion coefficient.
It is worth mentioning that the concept of Bloch wave function is originally
introduced for linear periodic systems \cite{58,59}, which can also be
extended to weakly nonlinear periodic quantum systems. Hence, it is
reasonably expected that the ground state of equation (\ref{eq5}) should occur to
the state of k = 0 in equation (\ref{eq6}).

An important feature of the Bloch state $\psi (x)$ in equation (\ref{eq6}) is that
the expansion coefficient $a_{m}$ in equation (\ref{eq7}) is \textit{complex} with
$a_{m}=|a_{m}|e^{i\phi _{m}}$. This feature immediately distinguishes the
Bloch state in an OSL from the one in an OL, where $a_{m}$ must be real. The
origin of such difference is that the conventional OL of $V_{\text{OL}}(x)$
has a parity symmetry ($V_{\text{OL}}(-x)=V_{\text{OL}}(x)$), which dictates
the imaginary part of $a_{m}$ to varnish, whereas an OSL of $V_{\text{OSL}%
}(x)$ usually does not have such symmetry, i.e. $V_{\text{OSL}}(-x)\neq V_{%
\text{OSL}}(x)$ for $\phi \neq 0$. In addition, the superposition of an
additional lattice gives rise to the emergence of substructures in each
lattice unit cell in the real space, which can result in substantial
modification of the band structure. For example, for suitable choices of $%
\phi $ and $V_{1(2)}$ as have been pointed out in references \cite{39,40,41,42}, the dispersion relation in the region between
the first and second Bloch band can be tuned to be linear, a feature
reminiscent of relativistic particles.

\subsection{Quantum fluctuations and regimes of validity of the mean-field theory}

For an optically trapped BEC, the phase fluctuations due to the periodic
potential will be inevitable to reduce the coherence of a BEC, leading to a
quantum phase transition from the superfluid to the Mott insulator. In this
work, we limit ourselves to the parameter regimes, where the mean-field
theory based on GP equation is always valid. Because the phase fluctuations
induced by OL depend on explicitly on the geometry of the model system, we
here emphasize the geometry of a BEC considered here as follows: along the $x
$-direction, the atoms can feel an OSL, in contrast, the model system is
uniform in the transverse direction. We now proceed to present two
approaches for justifying the validity of the mean-field theory.

Generally speaking, the mean-field theory based on GP equation is reliable
upon the quantum depletion of $N_{ex}$ being small relative to the total
atom number of $N$, i.e. $N_{ex}\ll N$. Hence, we posteriori justify the
mean-field theory by computing the quantum depletion of $N_{ex}/N$. After
that, we choose the safe parameter regimes by letting $N_{ex}/N\ll 1$.
Following reference \cite{19}, we can obtain the analytical expression of quantum
depletion of the condensate as follows,%
\begin{equation}
\frac{N_{ex}}{N}=2\frac{a_{3D}}{\sqrt{2\pi }\sigma }\left[ \frac{1}{2}-\frac{%
\sqrt{b}}{\pi }+\frac{b}{2}-\arctan \left( \sqrt{b}\right) \left( 1+b\right)
/\sqrt{\pi }\right], \label{eq8}
\end{equation}

with%
\begin{eqnarray}
b &=&\frac{2\sqrt{2}}{\sqrt{\pi ^{3}}}\frac{m}{m^{\ast }}\frac{\sigma }{d}%
\frac{E_{R}}{gn_{0}d}, \label{eq9}\\
\sigma  &=&\frac{d}{2\pi }\frac{1+1/16\sqrt{v_{1}+v_{2}\cos \left[ 2\phi %
\right] }}{\left( v_{1}+v_{2}\cos \left[ 2\phi \right] \right) ^{1/4}},
\label{10}
\end{eqnarray}

Based on equation (\ref{eq8}), two conclusions are immediately made: (i) equation
(\ref{eq8}) can recover the well-known 3D result of $N_{ex}/N=\left( 8/3\sqrt{\pi }%
\right) \sqrt{m^{\ast }/m}\left( na^{\prime3}\right) ^{1/2}$ with $%
a^{\prime}=a_{3D}d/\sqrt{2\pi }\sigma $ in teh limit of weak potential;
(ii) then, and the validity regimes of the mean-field theory can be further
identified by the conditions \cite{19} as%
\begin{equation*}
N_{ex}/N\approx m^{\ast }c_{s}d/\left( 2\pi \hbar N_{tot}\right) \log \left(
4N_{\perp }/\pi \right) \ll 1
\end{equation*}

with $N_{\perp }$ being the number of lattice wells along $x$-direction. As
emphasized in reference \cite{19}, the mean-field theory of equation (\ref{eq2}) is still
valid at $V_{1}=v_{1}\times 16E_{R}=20E_{R}$ under the conditions of the
typical experimental parameters $gn_{0}d=0.2E_{R}$\ and $N_{\perp }=200$.

On the other hand, we can justify the mean-field theory by checking the
asymptotical values of the off-diagonal one-body density matrix at large
distance as follows \cite{2,19}%
\begin{equation}
n^{\left( 1\right) }\left( \left\vert \mathbf{r}-\mathbf{r}^{^{\prime
}}\right\vert \right) \rightarrow \left\vert \mathbf{r}-\mathbf{r}^{^{\prime
}}\right\vert ^{-\gamma },
\label{eq11}
\end{equation}

with $\gamma =m^{\ast }c_{s}d/\left( 2\pi \hbar N_{well}\right) $. Here, $%
N_{well}$ is the number of particles per well. We have checked that the
exponent $\gamma $ is much smaller than 1 in our parameter regimes, which
means that the coherence survives at large distances and the application of
the mean-field theory based on GP equation is justified.

\section{Sound velocity}

\label{sec:Methods}

The previous section sets the stage for our study on the sound propagation
in a BEC trapped in an OSL. In this section, we outline both the analytical
and numerical approaches to the sound velocity of a BEC.

\subsection{Sound velocity in an optical lattice}

For an optically trapped BEC, the sound propagation and its velocity can be
viewed from two different perspectives \cite{26}. The first perspective
establishes the sound propagation in a BEC as a long-wavelength response to
an external perturbation, which thus connects the sound velocity to the
microscopic excitations of a BEC \cite{1,2}.
In an OSL, the sound velocity of a quasi-1D BEC is given by
\begin{equation}
c_{s}=\lim_{{p}\rightarrow 0}\frac{\epsilon _{p}}{\hbar p},
\label{eq12}
\end{equation}%
where $p$ and $\epsilon _{p}$ are the quasi-momentum of the probe and the
energy of the excitation, respectively. In equation (\ref{eq12}), the
effects of the lattice is encoded in the energy spectrum $\epsilon _{p}$.

In the second perspective, on the other hand, the sound velocity $c_{s}$ is
intimately related to the superfluidity of a BEC and its macroscopic
dynamics, where the sound velocity can be written as \cite{3,4,26}
\begin{equation}
c_{s}=\sqrt{\frac{1}{\kappa m^{\ast }}}.  \label{eq13}
\end{equation}%
Here, the compressibility $\kappa $ and the effective mass $m^{\ast }$, are
defined as, respectively,
\begin{equation}
\frac{1}{m^{\ast }}=\lim_{k\rightarrow 0}\frac{d^{2}E_{k}}{\hbar ^{2}dk^{2}},
\label{eq14}
\end{equation}%
and
\begin{equation}
\kappa ^{-1}=n_{0}\frac{\partial \mu }{\partial n_{0}},  \label{eq15}
\end{equation}%
with $n_{0}$ being the 3D condensate density, $E_{k}$ the average energy of
the system obtained by plugging equation (\ref{eq6}) into equation (\ref{eq3}) and $\mu
=\partial E_{k}/\partial n_{0}$ the chemical potential. The effect of an OSL
is thus encoded in the increased effective mass $m^{\ast }$ along the
lattice direction, and the enhanced interatomic interaction which manifests
itself in the decreased compressibility $\kappa $. The two definitions on
the sound velocity (Eqs. (\ref{eq12}) and (\ref{eq13})) have
been proved equivalent \cite{3,4,19} for an optically trapped BEC.

\subsection{Sound velocity in an OSL: perturbation approach}

We now analyze the effect of an OSL\ of\ $V_{\text{OSL}}(x)$ ($V_{2}\neq 0$
in Eq. (\ref{eq4})) on the sound velocity of a quasi-1D BEC. In this
section, we will assume the strengh of $v_{2}$ is weak, and then consider
two regimes associated with the strengh of $v_{1}$: in the first regime,
hereafter referred to as the weak superlattice regime, $v_{1}$ is comparable
with $v_{2}$ and are both small (say, $v_{1}\sim v_{2}\sim \lambda $ with $%
\lambda $ being introduced to label the order of perturbation and finally
set to be one); in the second regime, hereafter referred to as the
tight-binding regime, $v_{1}$ is much stronger than $v_{2}$, i.e. $v_{1}\gg
v_{2}$ (we assume $v_{1}$ is still sufficient to maintain the coherence of
BEC over several optical wells). Subsequently, we will derive analytically
the sound velocity in the two regimes, respectively.

\subsubsection{Weak superlattice regime}

In the weak superlattice regime with $v_{1}\sim v_{2}\sim \lambda $, the
whole superlattice potential $V_{\text{OSL}}(x)$ in Hamiltonian (\ref{eq3})
can be treated as a perturbation to a unperturbed system consisting of a
homogeneous quasi-1D BEC. In this case, we solve the GPE (\ref{eq5}) by
developing a perturbation expansion to the condensate wave function $\psi (x)
$ up to the second order of the small parameters $v_{1}\sim v_{2}\sim
\lambda $, i.e.
\begin{equation}
\psi (x)=\psi ^{(0)}(x)+\psi ^{(1)}(x)+\psi ^{(2)}(x)+o(\lambda ^{3}).
\label{eq16}
\end{equation}%
Direct application of the perturbation procedures \cite{26} to the GPE (%
\ref{eq5}) order by order thus yields $\psi ^{(0)}(x)=1$, as it is should
be, and
\begin{eqnarray}
\psi ^{(1)} &=&-\frac{1+k}{1+c-k^{2}}\frac{v_{2}}{4}e^{-2ix-i\phi }-\frac{%
2k-1}{4k^{2}-4c-1}v_{1}e^{-ix}  \notag \\
&+&\frac{2k+1}{4k^{2}-4c-1}v_{1}e^{ix}-\frac{1-k}{1+c-k^{2}}\frac{v_{2}}{4}%
e^{2ix+i\phi },  \label{eq17}
\end{eqnarray}%
as well as
\begin{eqnarray}
\psi ^{(2)} &=&\left[ \frac{2+2k}{4+4c-4k^{2}}C+\frac{c}{4+4c-4k^{2}}\left(
D-C\right) \right] v_{1}^{2}e^{-2ix}  \notag \\
&-&\left[ \frac{2\left( 1+2k+2c\right) }{1+4c-4k^{2}}B^{\ast }+\frac{4c}{%
1+4c-4k^{2}}A^{\ast }\right] v_{1}v_{2}e^{-ix}  \notag \\
&+&\left[ \frac{2\left( 1-2k+2c\right) }{1+4c-4k^{2}}A-\frac{4c}{1+4c-4k^{2}}%
B\right] v_{1}v_{2}e^{ix}  \notag \\
&+&\left[ \frac{(2-2k)C}{4+4c-4k^{2}}+\frac{c\left( C-D\right) }{4+4c-4k^{2}}%
\right] v_{1}^{2}e^{2ix}.  \label{eq18}
\end{eqnarray}%
Here, the coefficients $A$, $B$, $C$ and $D$ are given in the Appendix \ref%
{App}.

Substituting the condensate function $\psi (x)\approx \psi ^{(0)}(x)+\psi
^{(1)}(x)+\psi ^{(2)}(x)$ into the Hamiltonian (\ref{eq3}), we derive the
energy of the BEC and calculate the effective mass $m^{\ast }$ and the
compressibility $\kappa $ from equations (\ref{eq14}) and (\ref{eq15}),
respectively. The results are
\begin{eqnarray}
\frac{1}{\kappa }=c &+&\frac{8c}{\left( 1+4c\right) ^{3}}v_{1}^{2}+\frac{c}{%
2\left( 1+c\right) ^{3}}v_{2}^{2}  \notag \\
&-&\frac{27c}{\left( 1+c\right) ^{3}\left( 1+4c\right) ^{4}}%
v_{1}^{2}v_{2}\cos \left( \phi \right) ,  \label{eq19}
\end{eqnarray}%
and
\begin{eqnarray}
\frac{1}{m^{\ast }}=1 &-&\frac{8}{\left( 1+4c\right) ^{2}}v_{1}^{2}-\frac{1}{%
2\left( 1+c\right) ^{2}}v_{2}^{2}  \notag \\
&-&\frac{15}{\left( 1+c\right) ^{2}\left( 1+4c\right) ^{3}}%
v_{1}^{2}v_{2}\cos \left( \phi \right) .  \label{eq20}
\end{eqnarray}%
The sound speed in equation (\ref{eq13}) is therefore readily derived as
\begin{eqnarray}
c_{s} &=&\sqrt{c}\left( 1-\frac{16c}{\left( 1+4c\right) ^{3}}v_{1}^{2}-\frac{%
1+2c}{4\left( 1+c\right) ^{3}}v_{2}^{2}\right)   \notag  \label{eq21} \\
&-&\frac{27+15(1+c)(1+4c)}{(1+c)^{3}(1+4c)^{4}}v_{1}^{2}v_{2}\cos (\phi ).
\end{eqnarray}

Equation (\ref{eq21}) is one of the key results of this paper. When $%
v_{2}=0$, equation (\ref{eq21}) can recover the corresponding result in
reference \cite{26} and agrees with the conclusion that the sound velocity of a BEC
in an OL of $v_{1}\cos x$ decreases monotonically with $v_{1}$, as it should
be. In the presence of $v_{2}$ and $\phi \neq 0$, the situation is very
different, where the sound velocity can instead increase with increasing $%
v_{2}$ for $\phi =\pi $. This is shown in Fig. \ref{Fig:3}d, which
reflects a greater influence from $\kappa $ than $m^{\ast }$ on the sound
velocity. Such behavior can be expected from analysis of equation (\ref{eq21}), where the last two terms in the brackets are definitely
negative, whereas the terms in the second line can be either positive or
negative depending on $\phi $. In particular, if we choose $\cos \phi <0$,
the last term in the brackets of equation (\ref{eq21}) appears at the second
order of $v_{2}$, whereas in comparison the last term of equation (\ref{eq21}) emerges at the first-order. Consequently, in the weak potential
limit the sound speed must first increase with increasing $v_{2}$. This
observation can be intuitively summarized within following picture: the
variation of relative phase $\phi $ (see Fig. \ref{Fig:1}b) can lead to a
more tightly squeezed condensate, which will greatly decrease the value of
the $\kappa $ and give rise to an increased sound velocity.

\subsubsection{Tight-binding regime}

We now turn to the tight-binding regime with $v_{2}\ll v_{1}$. In this case,
the lattice potential $v_{2}\cos (2x+\phi )$ can be considered as a
perturbation, while the unperturbed system consists of a quasi-1D BEC
tightly confined in an optical lattice $v_{1}\cos (x)$. The unperturbed
tightly confined BEC system can be well described using the tight-binding
model \cite{60}, where the condensate wavefunction $\psi (x)$ can be written as a
superposition form as \cite{5}
\begin{equation}
\psi (x)=\sum_{n}\varphi _{n}(x)\psi _{n},  \label{eq22}
\end{equation}%
where $n$ is the site number of the lattice $v_{1}\cos (x)$, $\varphi _{n}(x)
$ is the condensate wavefunction in the $n$th well with $\varphi
_{n}(x)=\varphi _{0}\left( x+nd\right) $ and $\psi _{n}$ is the
corresponding expansion coefficients. In the presence of additional lattice
potential $v_{2}\cos (2x+\phi )$, for $v_{2}\ll v_{1}$, we expect that the
effect of the $v_{2}$ lattice is to cause small modifications to the
condensate function $\psi (x)$ in (\ref{eq22}). We therefore use a
variational approach and take equation (\ref{eq22}) as a variational ansatz
for the condensate function $\psi (x)$ in the presence of the $V_{2}$
lattice, with $\psi _{n}$ being the variational parameters. Substituting the
variational ansatz (\ref{eq22}) into equation (\ref{eq3}), we recast the
Hamiltonian $H_{1D}$ as
\begin{equation}
H=-J\sum_{n}\left( \psi _{n}^{\ast }\psi _{n+1}+\psi _{n+1}^{\ast }\psi
_{n}\right) +\frac{U}{2}\sum_{n}|\psi _{n}|^{4},  \label{eq23}
\end{equation}%
with the hopping amplitude
\begin{equation}
J=-\frac{1}{(2\pi )}\int dx\left[ \frac{1}{2}\left( \partial _{x}\varphi
_{n}\cdot \partial _{x}\varphi _{n+1}\right) +\varphi _{n}V_{\text{OSL}%
}\varphi _{n+1}\right] ,  \label{eq24}
\end{equation}%
and the on-site interaction
\begin{equation}
U=\frac{c}{(2\pi )}\int dx\varphi _{n}^{4}.  \label{eq25}
\end{equation}%
It follows from equations (\ref{eq23})-(\ref{eq25}) that the effect of a
weak $V_{2}$ lattice is to modify the hopping amplitude $J$ and the on-site
interaction $U$.

The ground state of Hamiltonian (\ref{eq23}) is a constant wave function $%
\psi _{{n}}=1$. As a standard procedure \cite{26}, its excitation
energy is given by $\epsilon (q_{x})=2|\sin \left( q_{x}\pi \right) |\sqrt{%
2J_{x}U}$. We now calculate $m^{\ast }$\ and $\kappa $ in terms of $J$ and $U
$ \cite{5}, respectively. First, in order to derive the compressibility $\kappa $
in equation (\ref{eq15}), we use a Gaussian form $\varphi _{n}(x)=(\pi ^{1/4}\sigma
_{x}^{1/2})\exp [-x^{2}/2\sigma _{x}^{2}]$ for the condensate function $%
\varphi _{n}(x)$ in calculating the $U$ in equation (\ref{eq25}), where the
width $\sigma _{x}$ of the condensate function is to be determined
variationally. The result isgiven by

\begin{equation}
\kappa =\frac{1+16\left[ v_{1}+4v_{2}\cos \left( \phi \right) \right] ^{1/2}%
}{\sqrt{16\pi }c\left[ v_{1}+4v_{2}\cos \left( \phi \right) \right] ^{3/4}}.
\label{eq26}
\end{equation}

It follows from equation (\ref{eq26}) that the compressibility $\kappa $ in
the tight-binding limit decreases with $v_{2}$ in a non-exponential form,
similarly as the weak potential limit. Next, we derive the effective mass in
equation (\ref{eq14}) by using similar methods to calculate $J$ as in reference \cite{27}. The resulting expression is written as
\begin{eqnarray}
\frac{m}{m^{\ast }} &=&\frac{1}{4}\left[ (d/\sigma _{x})^{4}-2(d/\sigma
_{x})^{2}\right] \exp \left[ -(d/2\sigma _{x})^{2}\right]   \notag \\
&-&8\pi ^{2}v_{1}\exp \left[ -(d/2\sigma _{x})^{2}-\pi ^{2}(\sigma
_{x}/d)^{2}\right]   \notag \\
&&+8\pi ^{2}v_{2}\cos \left( \phi \right) \exp \left[ -(d/2\sigma
_{x})^{2}-4\pi ^{2}(\sigma _{x}/d)^{2}\right]   \notag \\
&-&\frac{16c\pi ^{2}d}{\sqrt{2\pi }\sigma _{x}}\exp \left[ -(\sqrt{3}d/2%
\sqrt{2}\sigma _{x})^{2}\right] .  \label{eq27}
\end{eqnarray}%
Note that equation (\ref{eq27}) is explicitly a single-particle result and is
only valid for weak interactions. Plugging equations (\ref{eq26}) and (\ref{eq27}) into
equation (\ref{eq13}), the analytical expression of sound velocity in the
tight-binding limit can be obtained, which is too complex to write down here.

We stress that the tight-binding model here is restricted to weak $V_{2}\ll
V_{1}$ as is shown in Figure \ref{Fig:1}b, where the substructure in each
well can be safely neglected. For strong $V_{2}$, the substructure has to be
taken into account and the tight-binding treatment in reference \cite{46} is a reliable method.

\subsection{Numerical methods in the whole regime}

\label{sec:Results}

In the previous section, we have considered the case where the intensity $%
V_{2}$ is weak, while the intensity $V_{1}$ is either small ($V_{1}\sim V_{2}
$) or much more stronger ($v_{1}\gg v_{2}$). Our analytical result already
reveals new features on the sound velocity caused by the superlattice, when
compared to the monochromatic optical lattice potential. In this section, we
will address the problem of calculating the sound velocity in a more broader
regimes using the numerical approach.

The key step of our approach consists in numerically solving the GPE
equation (\ref{eq5}) based on the Bloch expansion in equation (\ref{eq7}) for the
condensate function $\psi (x)$ (the expansion is cut off at a certain number
$m=N$ in the numerical calculations). After the Bloch expansion coefficients
$a_{m}$ are found and the condensate function $\psi (x)$ are obtained, we
accordingly calculate the energy $E_{k}$ from equation (\ref{eq3}), which
will lead to the derivation of the compressibility $\kappa $, the effective
mass $m^{\ast }$ and the sound velocity $c_{s}$ thereof. In order to
comprehensively reveal the effect of superlattice, including the strength of
the individual constituting lattice ($v_{1}$ and $v_{2}$), and the relative
phase $\phi $, on the sound velocity, we have considered three cases for
numerical analysis:
\begin{figure}[tbp]
\includegraphics[width=0.46\textwidth]{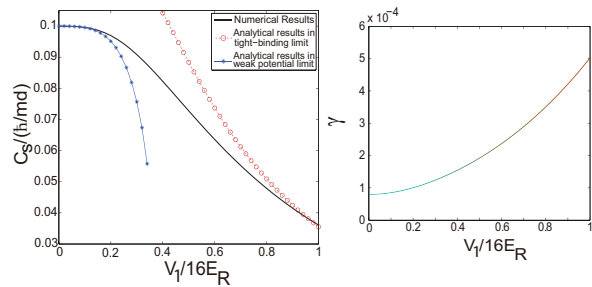}
\caption{(Color online) Left panel: sound velocity $c_{s}$ in unit of $\hbar/md$ for a BEC
in an OSL via the lattice strength of $v_{1}=V_{1}/16E_{R}$. The
numerical results are denoted by the solid lines, analytical results
from equation (\ref{eq21}) in weak potential limit by stars, and analytical tight-binding
results by circles obtained by plugging equations (\ref{eq26}) and (\ref{eq27}) into equation (\ref{eq13}).
Right panel: the exponent $\gamma$ of the off-diagonal one-body density
matrix defined in equation (\ref{eq11}) via the lattice strength of $v_{1}=V_{1}/16E_{R}$,
which justifies the validity of the mean-field theoty. The parameters
are given by $gn_{0}d=c\times8E_{R}$, $V_{2}=v_{2}\times16E_{R}=0.48E_{R}$,
and $\phi=2\theta+\pi=0$, respectively.}
\label{Fig:2}
\end{figure}

\begin{figure}[tb]
\includegraphics[width=0.46\textwidth]{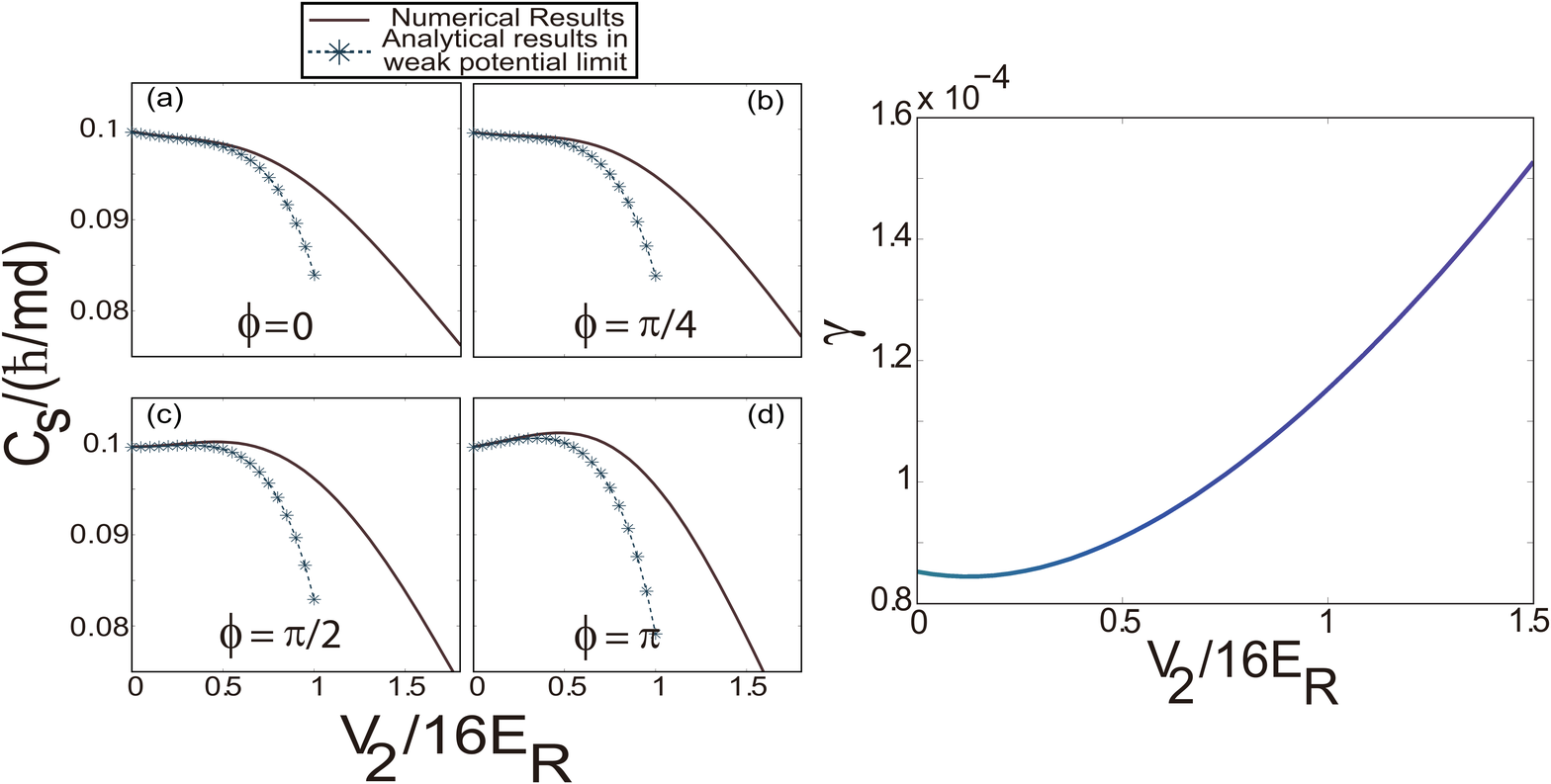}
\caption{(Color online) Left panel: sound velocity $c_{s}$ in unit of $\hbar/md$ for a BEC
in an OSL via the lattice strength of $v_{2}=V_{2}/16E_{R}$. The
numerical results are denoted by the solid lines and analytical results
from equation (\ref{eq21}) in weak potential limit by stars. (a) $\phi=2\theta+\pi=0$;
(b) $\phi=2\theta+\pi=\pi/4$; (c) $\phi=2\theta+\pi=\pi/2$; (d)
$\phi=2\theta+\pi=\pi$. Right panel: the exponent $\gamma$ of off-diagonal
one-body density matrix defined in equation (\ref{eq11}) via the lattice strength
of $v_{2}=V_{2}/16E_{R}$ with $\phi=2\theta+\pi=0$, which justifies
the validity of the mean-field theory. The parameters are given by
$gn_{0}d=c\times8E_{R}=0.08E_{R}$ and $V_{1}=v_{1}\times E_{R}=1.6E_{R}$,
respectively. }
\label{Fig:3}
\end{figure}

(i) In the first step (see Fig. \ref{Fig:2}), we fix the intensity $%
v_{2}=V_{2}/16E_{R}$ at a small value ($V_{2}\sim 0.5E_{R}$) as well as
fixing the relative phase $\phi $, and scan the sound velocity $c_{s}$ as a
function of $v_{1}=V_{1}/16E_{R}$. As is shown in the left panel of Figure
\ref{Fig:2}, the sound velocity $c_{s}$ decreases monotonically with increasing $V_{1}$
as expected. For small $V_{1}\sim V_{2}$ (weak potential limit) and for $%
V_{1}\gg V_{2}$ (tight-binding limit), we see that the numerical result
agrees well with our analytical results as they should be. Moreover, we have
checked that our numerical results in limit of $V_{2}=0$ can recover the
corresponding one in references \cite{19,26}.

(ii) In the second step (see Fig. \ref{Fig:3}), we fix $v_{1}=V_{1}/16E_{R}$
at a small value ($V_{1}\sim 1.6E_{R}$ in Figure \ref{Fig:3}), and scan the
sound velocity $c_{s}$ as a function of $v_{2}=V_{2}/16E_{R}$ for various
choice of $\phi $. As is shown, in the upper two figures of Figure \ref%
{Fig:3} ($\phi =0,\pi /4$, respectively), the sound velocity still shows an
overall decrease with increasing $V_{2}$. Whereas, when $\phi =\pi /2,\pi $,
it is clearly shown that the sound velocity $c_{s}$ firstly increases and
then decreases with increasing $V_{2}$. The numerical results depicted in
Figure \ref{Fig:3} can be understood using equation (\ref{eq21}): the
last term in equation (\ref{eq21}) can be positive or negative depending
on $\phi $, whereas the last two terms in the bracket of equation (\ref{eq21}) are always negative. Naively, one would expect that the sound
velocity either decreases monotonically with $V_{2}$ for the case of $\cos
\phi >0$ (see Figs. \ref{Fig:3}a and \ref{Fig:3}b) or may develop a maximum for $\cos \phi <0$
(see Figs. \ref{Fig:3}c and \ref{Fig:3}d). This is exactly what we have seen in Figure \ref%
{Fig:3}. However, for large $V_{2}$ in Fig. \ref{Fig:3}, a marked
discrepancy emerges between the tight-binding results and the numerical one.
This can be traced back to the fact that the substructures in each lattice
well have been neglected in the tight-binding treatment, which nevertheless
plays increasingly important role on determining the sound speed for larger $%
V_{2}$.

\begin{figure}[tb]
\includegraphics[width=0.46\textwidth]{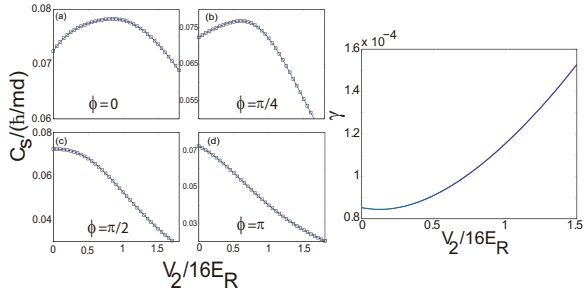}
\caption{(Color online) Left panel: sound velocity $c_{s}$ in unit of $\hbar/md$ for a BEC
in an OSL via the OSL strength of $v_{2}=V_{2}/16E_{R}$. (a) $\phi=2\theta+\pi=0$;
(b) $\phi=2\theta+\pi=\pi/4$; (c) $\phi=2\theta+\pi=\pi/2$; (d)
$\phi=2\theta+\pi=\pi$. Right panel: the exponent $\gamma$ of off-diagonal
one-body density matrix defined in equation (\ref{eq11}) via the OSL strength
of $v_{2}=V_{2}/16E_{R}$, which justifies the validity of the mean-field
theory. The parameters are given by $gn_{0}d=c\times8E_{R}=0.08E_{R}$
and $V_{1}=v_{1}\times16E_{R}=1.6E_{R}$, respectively.}
\label{Fig:4}
\end{figure}

\begin{figure}[tbp]
\includegraphics[width=0.46\textwidth]{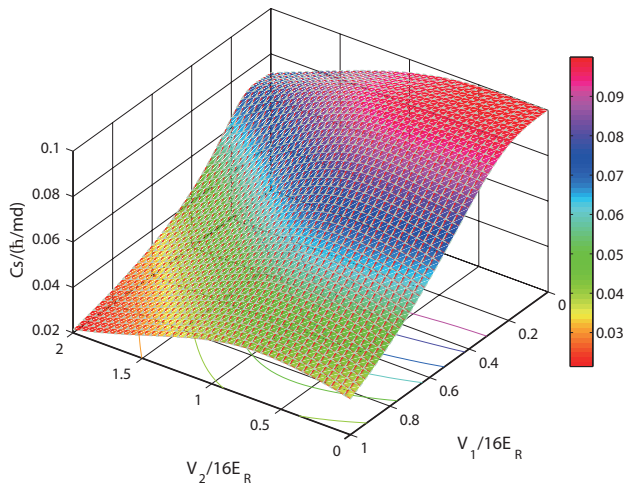}
\caption{(Color online) Sound velocity $c_{s}$ in unit of $\hbar/md$ for a BEC in an OSL
via both the lattice strengths of $v_{1}=V_{1}/16E_{R}$ and $v_{2}=V_{2}/16E_{R}$.
The parameters are given by $gn_{0}d=c\times8E_{R}=0.08E_{R}$ and
$\phi=2\theta+\pi=0$.}
\label{Fig:5}
\end{figure}

(iii) To further highlight the interplay between the two lattices with
different periods, in the third step (see Fig. \ref{Fig:4}), we choose a
strong intensity $v_{1}=V_{1}/16E_{R}$ ($V_{1}=9.6E_{R}$ in Fig. \ref{Fig:4}%
), and scan the sound velocity as a function of $V_{2}$ for various choice
of $\phi $. As shown in Figures \ref{Fig:4}a and \ref{Fig:4}b,  it turns out
that the sound velocity achieves a maximum value for $\phi =0$, $\pi /4$.
These results in Figures \ref{Fig:4}a and \ref{Fig:4}b are not conflict with
equation (\ref{eq21}), which suggests that a maximum of sound velocity can be only
developed for $\phi =0$, $\pi /4$, because equation (\ref{eq21}) becomes to be
invalid in the case of strong $V_{1}$. We attribute the explanation for
Figures \ref{Fig:4}a and \ref{Fig:4}b to the competition between the
compressibility and effective mass through equation (\ref{eq13}) and such
competition is supposed to manifest itself quite differently with respect to
the various choices of OSL's parameters, resulting in the difference between
Figures \ref{Fig:3} and \ref{Fig:4}. On the other hand, motivated by
reference \cite{61}, another possible intuitive explanation for sound velocity
in\ Figures \ref{Fig:3} and \ref{Fig:4}\ can be carried out in terms of the
effective dimensionality seen by a BEC. The lattice of $V_{2}$ consists of
the $x$-direction of our system, while the introduction of the strong
strength $V_{1}$ plays the role of squeezing a BEC reminiscent of the other
spatial dimensionality of the model system. Therefore, we can reasonably
view our system as an effective 2D BEC loaded in a monochromatic OL with the
role of the dimensionality taken by the $V_{1}$ from the view of they both
squeezing the condensate. Figures \ref{Fig:3} and \ref{Fig:4} become to be
understood in the framework of reference \cite{26}, i.e. sound velocity of an
optically-trapped BEC in 2D may develop a maximum with the increasing
lattice strength. More importantly, in Figure \ref{Fig:4}, the maximum of
sound velocity can exceed that for $V_{2}=0$ with the same $V_{1}$ by $10\%$%
. Moreover, the maximum falls into the superfluid regimes, and should be
observable within the current experimental conditions. Finally, to make our
investigation of sound velocity more completely, we have plotted Figure \ref{Fig:5}
with the fixed value of $\phi =0$, which shows how the sound velocity varies
via both the $V_{1}/16E_{R}$ and $V_{2}/E_{R}$.

Meanwhile, as shown by the right panel of Figures \ref{Fig:2}-\ref{Fig:4}, we have monitored
that the exponent $\gamma$ of off-diagonal one-body density matrix is
always much smaller than 1, justifying the calculations of sound speed in
the framework of the mean-field theory based on GPE. Note that our study on
the sound propagation has been done based on the GPE which consists of
ignoring the quantum fluctuations and temperature effects. To study the
effects of finite temperature and fluctuations, particularly near the
transition point of superfluid and Mott insulator, one has to use other
theories \cite{53}.

\section{Possible experimental scenarios and conclusion}

\label{sec:Summary}

The present model on the sound velocity of a quasi-1D BEC in an OSL is based
on GPE and highlights the competition between the compressibility $\kappa $
and the effective mass $m^{\ast }$. We have shown that the sound velocity is
characterized by four parameters: the effective interatomic interaction $c$,
the lattice intensities $V_{1(2)}$ and the relative phase $\phi $ between
the fundamental lattice and the double-period lattice. All these quantities
are experimentally controllable using state-of-the-art technologies. The
interatomic interaction can be controlled in a very versatile manner via the
technology of Feshbach resonances \cite{62}. In typical experiments to
date \cite{5}, the values of $c$ range from $0.01$ to $0.1E_{R}$. The
strength of both $V_{1}$ and $V_{2}$ an be tuned from $0E_{R}$ to $32E_{R}$
almost at will. Phase control \cite{46} between the two standing wave
fields allows $\phi $ to scan the whole range $[0,2\pi ]$.

Central to testing the validity of the prediction in this article is the
experimental ability to measure the sound velocity in a BEC. The
experimental approaches so far to sound velocity are based on analyzing the
linear response of a fluid to an external velocity boost. The key quantity
to measure in these approaches is the dynamic structure factor of the model
system \cite{3,4}. The speed of sound of a BEC in an OL may be measured
with a similar technique as was used in references \cite{9,63}. Another option is to employ Bragg spectroscopy \cite{10,11,64} to the excitation spectrum. The sound velocity can be
extracted from the slope of the linear part of the excitation spectrum.

Another difficulty may arise from the typical application of an external
harmonic trap with the frequency of $\omega $ in the BEC experiments, which
will lead to the discrete spectrum due to the finite size of the trapped
BEC \cite{65}. The detailed concern is as follows: our definition of sound velocity is
based on equation (\ref{eq13}), where the presence of the OL has been
accounted for through renormalizing compressibility $\kappa $ and the
effective mass $m^{\ast }$. As a result, despite the presence of the
lattice, one can study the sound velocity of a BEC using equation (\ref{eq13}) as if the space is homogeneous. In such, with the
considerations of the harmonic trap, the condensate wave-function is
non-vanishing over a size of the $l=(\sqrt{2}/\pi )c_{s}/\omega $ in the
Thomas-Fermi approximation. As a compression, the frequencies of the lowest
collective excitations are proportional to $c_{s}/l\propto \omega $, being
simply proportional to the trapping frequencies, which do not yield direct
information of the sound velocity. Therefore, in order to observe the sound
speed, one must excite the perturbation much smaller than the size of the
condensate by ruling out the collective excitation of the model system.

On overcoming the preceding two difficulties, the experimental realization
of our scenario amounts to controlling four parameters whose interplay
underlies the physics of this work. Therefore, the phenomena discussed in
this article should be observable within the current experimental
capability, which would constitute an important step in understanding the
effect of an OSL on the superfluidity of a BEC.

In summary, we have studied analytically and numerically the sound speed of
a 1D BEC in an OSL. Our results show that the interplay between two
constituting lattices that have different periods can significantly
influence the sound velocity of the BEC system. In particular, unusual
behavior of sound propagation in an OSL compared to the case in an OL at 1D
has been found, i.e. the sound speed can first increase and then decrease as
the lattice strength $V_{2}$ increases. Such behavior can be understood
using our analytical results for weak lattices. As a consequence, an
experimentalist can, in principle, engineer the rich behavior of the sound
speed by altering OSL's parameters, which can find the direct applications
for manipulating system bath coupling in driven open system with atoms.

\section*{Acknowledgments}

We thank Ying Hu and Biao Wu for helpful discussions and Teng Yang for
carefully reading our manuscript. This work is supported by the NSF of China
(Grants Nos. 11004200 and 11274315).

\appendix

\section{Preliminary notations}

\label{App}

The explicit expressions of $A$, $B$, $C$, and $D$ in equation (\ref{eq18}%
) read as follows respectively,
\begin{eqnarray}
A &=&\frac{1}{8}\frac{\left( 4k^{2}-7\right) \left( k+2c+2\right) +9\left(
2c+1\right) }{\left( 1+c-k^{2}\right) \left( 4k^{2}-4c-1\right) }e^{i\phi },
\\
B &=&-\frac{1}{8}\frac{\left( 4k^{2}-7\right) \left( k-2c-2\right) -9\left(
2c+1\right) }{\left( 1+c-k^{2}\right) \left( 4k^{2}-4c-1\right) }e^{i\phi },
\\
C &=&\frac{1}{2}\frac{\left( 4k^{2}-1\right) \left( 2k+2c-1\right) }{\left(
4k^{2}-4c-1\right) ^{2}}, \\
D &=&-\frac{1}{2}\frac{\left( 4k^{2}-1\right) \left( 2k-2c+1\right) }{\left(
4k^{2}-4c-1\right) ^{2}}.
\end{eqnarray}

\end{document}